\documentclass[a4paper,12pt]{article}
\pdfoutput=1 

\usepackage{jheppub} 

\usepackage[T1]{fontenc} 
\usepackage[utf8]{inputenc}
\usepackage{amsmath}
\usepackage{xspace}

\newcommand{\meaneps}{\langle{\epsilon}\rangle}
\newcommand{\be}{\begin{equation}}
\newcommand{\ee}{\end{equation}}
\newcommand{\ben}{\begin{equation*}}
\newcommand{\een}{\end{equation*}}
\newcommand{\bea}{\begin{eqnarray}}
\newcommand{\eea}{\end{eqnarray}}
\newcommand{\bean}{\begin{eqnarray*}}
\newcommand{\eean}{\end{eqnarray*}}
\newcommand{\bi}{\begin{itemize}}
\newcommand{\ei}{\end{itemize}}
\newcommand{\dd}{{\rm d}}
\def\cO#1{{{\cal{O}}}\left(#1\right)}

\newcommand{\rpa}{\ensuremath{R_{\rm pA}}\xspace}

\newcommand{\rpia}{\ensuremath{R_{\rm {\pi}A}}\xspace}

\newcommand{\tf}{\ensuremath{t_{\mathrm{f}}}\xspace}
\newcommand{\Ebeam}{\ensuremath{E_\mathrm{b}}\xspace}
\newcommand{\xf}{\ensuremath{x_{\mathrm{F}}}\xspace}
\newcommand{\qzero}{\hat{q}_0}
\newcommand{\gevsqfm}{GeV$^2$/fm\xspace}
\newcommand{\qhat}{\hat{q}}
\def\X{{\rm X}}
\def\xone{\ensuremath{x_{1}}\xspace}
\def\xtwo{\ensuremath{x_{2}}\xspace}
\newcommand{\ie}{{i.e.}\xspace}
\newcommand{\eg}{{e.g.}\xspace}
\newcommand{\lqcd}{\Lambda_{_{\rm QCD}}} 
\newcommand{\alphas}{\alpha_s}

\newcommand{\A}{{\rm A}}
\def\cO#1{{{\cal{O}}}\left(#1\right)}
\newcommand\Pdist{{\cal P}}
\newcommand{\Jpsi}{\ensuremath{{J}\hspace{-.08em}/\hspace{-.14em}\psi}\xspace}
\newcommand{\jpsi}{\Jpsi}
\newcommand{\B}{\ensuremath{{\text{B}}}\xspace}
\newcommand{\eq}[1]{Eq.~\eqref{#1}\xspace}
\newcommand{\pp}{{\ensuremath{pp}}\xspace}
\newcommand{\pA}{\ensuremath{\text{pA}}\xspace}
\newcommand{\hone}{\ensuremath{\text{h}_1}\xspace}
\newcommand{\htwo}{\ensuremath{\text{h}_2}\xspace}
\newcommand{\hh}{\ensuremath{\text{h}_1\text{h}_2}\xspace}
\newcommand{\hA}{\ensuremath{\text{hA}}\xspace}
\newcommand{\hB}{\ensuremath{\text{hB}}\xspace}
\newcommand{\piA}{\ensuremath{\pi\text{A}}\xspace}
\newcommand{\sqrts}{\ensuremath{\sqrt{s}}\xspace}

\DeclareFontFamily{T1}{calligra}{}
\DeclareFontShape{T1}{calligra}{m}{n}{<->s*[1.44]callig15}{}
\DeclareMathAlphabet\mathcalligra   {T1}{calligra} {m} {n}
\DeclareMathAlphabet\mathzapf       {T1}{pzc} {mb} {it}
\DeclareMathAlphabet\mathchorus     {T1}{qzc} {m} {n}
\DeclareMathAlphabet\mathrsfso      {U}{rsfso}{m}{n}
\title{\boldmath Initial-state energy loss in cold QCD matter and the Drell-Yan process}
\usepackage{float}
\usepackage{braket}
\renewcommand*{\thefootnote}{\fnsymbol{footnote}}

\author[a]{François Arleo}
\author[a,b]{\hspace{-0.2cm}, Charles-Joseph Naïm\footnote{corresponding author}}
\author[b]{\hspace{-0.2cm}, Stephane Platchkov}

\affiliation[a]{Laboratoire Leprince-Ringuet, École polytechnique, CNRS/IN2P3,  Université Paris-Saclay,  91128, Palaiseau, France}
\affiliation[b]{IRFU, CEA, Université Paris-Saclay, 91191 Gif-sur-Yvette, France}
\emailAdd{francois.arleo@cern.ch}
\emailAdd{charles-joseph.naim@cern.ch}
\emailAdd{stephane.platchkov@cern.ch}

\abstract{The effects of parton energy loss in nuclear matter on the Drell-Yan process in pA and $\pi$A collisions at fixed-target energies are investigated. Calculations are based on the Baier-Dokshitzer-Mueller-Peigné-Schiff (BDMPS) framework embedded in a next-to-leading order calculation, using the transport coefficient extracted from $J/\psi$ measurements. Model calculations prove in good agreement with preliminary measurements by the E906 experiment, despite a slightly different magnitude, supporting a consistent picture between Drell-Yan and $J/\psi$ data. Predictions for the COMPASS future measurements in $\pi$A collisions at $\sqrt{s}=18.9$~GeV are also performed. At higher collision energy ($\sqrt{s}=38.7$~GeV), Drell-Yan measurements are only slightly affected by energy loss effects. On the contrary, the E906 results turn out in clear disagreement with nuclear PDF effects alone. The comparison of E772, E866, and E906 measurements indicates for the first time a clear violation of QCD factorization in Drell-Yan production in pA collisions.}
\begin{document}
\maketitle
\flushbottom

\renewcommand*{\thefootnote}{\arabic{footnote}}
\setcounter{footnote}{0}

\section{Introduction}
\label{sec:intro}
The celebrated jet quenching phenomenon, observed in heavy ion collisions at RHIC and LHC, indicates that quarks and gluons experience radiative energy loss while propagating in a hot, deconfined QCD medium (see Refs.~\cite{Majumder:2010qh,Mehtar-Tani:2013pia,Armesto:2015ioy,Qin:2015srf} for recent reviews). The wealth of data collected so far has triggered detailed phenomenological studies serving a dual purpose, namely (i) to probe the radiative energy loss of partons (and multipartonic states) in a medium, and (ii) to extract the scattering properties of the expanding medium produced in these collisions, eventually leading to a better understanding of hot QCD matter.

Another way to study parton energy loss is to consider the production of hard QCD processes in hadron-nucleus (hA) collisions; see Ref.~\cite{Arleo:2016lvg} for a recent discussion. In this case the medium, cold nuclear matter, is simpler: it is static with known size and nuclear density. It is not less interesting, though, as it reveals important features of medium-induced gluon radiation expected in QCD. More explicitly, hard processes in hA collisions are sensitive to different timescales of gluon radiation: the Landau-Pomeranchuk-Migdal (LPM) regime, corresponding to gluon formation times of the order of the medium length ($\tf \lesssim L$), and the factorization regime (also known as {\it fully coherent}) for which $\tf \gg L$~\cite{Peigne:2008wu}.

The latter, fully coherent radiative energy loss, arises from the interference of gluon emission amplitudes off an `asymptotic' incoming particle, produced long before entering the medium, and an asymptotic outgoing particle~\cite{Arleo:2010rb,Arleo:2012hn,Arleo:2012rs,Peigne:2014uha,Peigne:2014rka}. It thus differs from the initial-state (final-state) energy loss of a given incoming (outgoing) particle. In this regime, the average energy loss associated to the production of a massive particle (mass $M$) scales \emph{linearly} with the particle energy $E$ in the rest frame of the medium,~\cite{Peigne:2014uha}
\begin{equation}\label{eq:coherent}
    \meaneps_{\rm coh} \propto (C_R + C_{R^\prime} - C_t) \cdot \frac{\sqrt{\qhat L}}{M}\cdot E\,,
\end{equation} 
where $C_R$, $C_{R^\prime}$ and $C_t$ are respectively the color charge (Casimir) of the incoming, outgoing and exchanged\footnote{between the projectile hadron and the target nucleus} particle ($C_R=4/3$ for a quark and $C_R=3$ for a gluon), and $\qhat$ is the transport coefficient of cold nuclear matter.
The fully coherent energy loss could play a key role in the suppression of hard processes in hA collisions. It was shown in particular that this sole process is able to reproduce $\jpsi$ suppression data in hA collisions~\cite{Arleo:2012rs,Arleo:2013zua}, from SPS to LHC energy and on a wide range in Feynman-$x$ (\xf), in contradistinction to other  effects such as nuclear modifications of parton distribution functions.

On the contrary, initial-state (or, final-state) energy loss is only sensitive to the LPM regime~\cite{Peigne:2014uha}, for which the average energy loss is independent of the parton energy (up to a logarithmic dependence),~\cite{Peigne:2008wu}

\begin{equation}\label{eq:LPM}
    \meaneps_{\rm LPM} \propto C_R \cdot \qhat L^2
\end{equation}
where now $C_R$ is the Casimir of the propagating particle. The fact that the LPM fractional energy loss, $\meaneps_{\rm LPM}/E$, vanishes in the high energy limit has important consequences for the phenomenology. In particular, the effects of initial-state (or, final-state) energy loss in nuclear matter should be negligible in hA collisions at high energy, $\meaneps_{\rm LPM} \ll E$, as the particle energy \emph{in the nucleus rest frame} gets large, $E \propto \sqrts$.

Unless the particle energy becomes very small, $E \lesssim M \sqrt{\qhat L^3}$, the fully coherent energy loss exceeds that in the LPM regime, $\meaneps_{\rm coh} \gg \meaneps_{\rm LPM}$. However, not all processes are sensitive to fully coherent energy loss. One such case is particle production at large angle\footnote{At large angle, gluon emissions along the incoming parton and along the outgoing parton would not overlap, thus suppressing interferences.} (with respect to the beam axis) in the medium rest frame: this typically corresponds to the case of jet quenching in heavy ion collisions. Consequently, an energetic parton propagating in a hot medium is expected to experience final-state energy loss according to the LPM regime, \eq{eq:LPM}. On the contrary, a particle produced in hA collisions is almost always produced with a large rapidity in the nucleus rest frame, thus at a `small' angle, making it potentially sensitive to fully coherent energy loss~\cite{Arleo:2010rb}.
Another process should be insensitive to fully coherent energy loss, namely the single inclusive production of Drell-Yan (DY) lepton pairs, since at leading order the final state is color neutral and therefore does not radiate gluons.\footnote{This can be seen from \eq{eq:coherent}: in the DY case, $C_{R^\prime}=0$ and $C_R=C_t$, leading to a vanishing color prefactor.} Hence, the production of DY pairs in hA collisions appears as a promising candidate in order to probe LPM initial-state energy loss in cold nuclear matter.\footnote{At high collision energy, \eg at RHIC or LHC, LPM initial-state energy loss in nuclear matter is negligible, making DY production in \pA collisions an ideal process to probe nuclear parton densities~\cite{Arleo:2015qiv}.} At next-to-leading order (NLO), the production of a virtual photon in association with a hard parton in the final state would make DY production potentially sensitive to fully coherent radiation; however, the dominant NLO subprocess at large \xf is Compton scattering, $qg \to q \gamma^\star$, for which the fully coherent radiation is small $(\propto 1/N_c)$ and negative.\footnote{In addition, these small effects could be balanced by the real NLO annihilation process, $q\bar{q}\to\gamma^\star g$, significantly suppressed compared to Compton scattering~\cite{Vogt:1999dw} but more sensitive to fully coherent radiation ($\propto N_c$) and positive~\cite{Peigne:2014uha}.} Final-state energy loss in nuclei could also be probed from the measurements of hadron production in semi-inclusive deep inelastic scattering (SIDIS) events~\cite{Wang:2002ri,Arleo:2003jz}. These processes are summarized in Table~\ref{tab:energylosses}.

{\footnotesize
\begin{table}[h]
    \centering
    \begin{tabular}[c]{p{2.7cm}p{5.8cm}p{2.8cm}c}
    \hline
    \hline
    Energy loss    & Process &  Regime & $\meaneps$ \\
    \hline
    Initial-state   & $\text{h}\A\to \ell^+\ell^-+\X$ (LO Drell-Yan)  &  $\tf \lesssim L$ (LPM) & $\propto\qhat L^2$ \\[0.25cm]
    Final-state   & $e\A\to e + h + \X$ (SIDIS)  &  $\tf \lesssim L$ (LPM) & $\propto\qhat L^2$ \\[0.125cm]
    \hline \\[-0.25cm]
    Fully coherent & $\text{h}\A\to [Q\bar{Q}(g)]_8 + \X$  &  $\tf \gg L$ (fact.)& $\propto\sqrt{\qhat L}/M\cdot E$ \\
    & \hfill (quarkonium)  & & \\
    \hline
    \hline
    \end{tabular}
    \caption{Probing energy loss in cold nuclear matter in different processes.}
    \label{tab:energylosses}
\end{table}}

Until now, the interpretation of DY data in hA collisions has been delicate and ambiguous. The E772 and later the E866 experiment at FNAL performed high-statistics measurements of DY pairs in \pA collisions at $\sqrts=38.7$~GeV, on a wide range in \xf, $0.1\lesssim \xf \lesssim 0.9$ ~\cite{Alde:1990im,Vasilev:1999fa}. The depletion observed in heavier nuclei (Fe, W) at large $x_F$ could either be attributed to nuclear parton distribution (nPDF) effects, namely sea quark shadowing at $x_2\gtrsim 10^{-2}$~\cite{deFlorian:2011fp,Kovarik:2015cma,Eskola:2016oht} or to strong energy loss effects in cold nuclear matter~\cite{Johnson:2001xfa,Neufeld:2010dz,Song:2017wuh}, therefore preventing a clear interpretation of these data.
Older and less precise NA3 data in \piA collisions at $\sqrts=16.8$~GeV~\cite{Badier:1981ci} proved less sensitive to nPDF effects and allowed for setting upper limits on parton energy loss in nuclei; however these measurements were also compatible with vanishing parton energy loss~\cite{Arleo:2002ph}. Therefore, because of both the poorly known sea quark shadowing and the large experimental uncertainties in earlier data, no clear evidence for parton energy loss in the Drell-Yan process has yet been found. Two fixed-target experiments now make it possible to better understand the origin of the DY nuclear dependence. The E906 experiment~\cite{E906} recently performed preliminary measurements of DY production in \pA collisions at $\sqrts=15$~GeV on a wide range of \xf~\cite{Lin:2017eoc}. In addition, the COMPASS experiment at the SPS collected data in \piA collisions at $\sqrts=18.9$~GeV that could be used to determine DY nuclear production ratios also on a large \xf interval~\cite{Aghasyan:2017jop}.

The goal of this article is to revisit the effects of LPM initial-state energy loss in the BDMPS formalism on DY production in hA collisions at fixed target energies ($\sqrts < 40$~GeV), with a systematic comparison between model calculations and experimental results. The theoretical framework is presented in Section~\ref{sec:model} and results are shown in Section~\ref{sec:phenomenology}. The violation of QCD factorization in Drell-Yan production in \pA collisions is discussed in Section~\ref{sec:x2scaling}. Conclusions are drawn in Section~\ref{sec:summary}.

\section{Drell-Yan production in hA collisions}\label{sec:model}

\subsection{NLO production cross section}

We consider the inclusive production of Drell-Yan lepton pairs of large invariant mass, $M \gg \lqcd$, in hadronic collisions.  The analysis is carried out at next-to-leading order accuracy in the strong coupling constant, \ie at order  $\cO{\alpha^2\,\alphas}$, using the DYNNLO Monte Carlo program~\cite{Catani:2007vq,Catani:2009sm}. The \xf-differential production cross section in a generic $\hh$ collision reads
\begin{equation}
    \label{eq:DYxs}
     \frac{\dd\sigma(\hh)}{\dd\xf\,\dd M} =   \sum\limits_{i,j=q,\bar{q},g} \int_{0}^{1} \, \dd x_1 \int_{0}^{1}\, \dd x_2\, f_{i}^{\hone}(x_{1}, \mu_R^2) f_{j}^{\htwo}(x_{2}, \mu_R^2)\, \frac{\dd\widehat{\sigma}_{ij}}{\dd\xf\,\dd M}(x_1 x_2 s, \mu^2, \mu_R^2)\,.
\end{equation}
At NLO, the partonic cross section $\hat{\sigma}_{ij}$ includes Compton scattering and annihilation processes, $qg \to q \gamma^{\star}$ and $q\bar{q} \to g \gamma^{\star}$, in addition to virtual corrections to the Born diagram, $q\bar{q}\to \gamma^\star$. In \eq{eq:DYxs}, both the renormalization and factorization scales are set equal to the DY invariant mass,\footnote{In the following, the explicit scale dependence is omitted for clarity.} $\mu_{R}^{2} = \mu^{2} = M^2$. The single differential cross section $\dd\sigma/\dd\xf$ is obtained by integrating \eqref{eq:DYxs} over the dilepton mass range, here between the charmonium and bottomonium resonances.

In this analysis we are interested in the production of Drell-Yan pairs using either a proton or a pion beam on nuclear targets, \pA and \piA collisions. In the absence of genuine nuclear effects, $f_j^{\htwo}$ appearing in \eq{eq:DYxs} should thus be replaced by the corresponding average over proton and neutron partonic densities,
\begin{equation}\label{eq:PDFnucleus}
    f_j^{\A}(x_{2}) = Z\,f_{j}^{p}(x_{2}) + (A-Z)\,f_{j}^{n}(x_{2})
\end{equation}
where $Z$ and $A$ are respectively the atomic and the mass number of the nucleus A.
Several proton NLO PDF sets have been used in this analysis (MMHT2014~\cite{Harland-Lang:2014zoa}, nCTEQ~\cite{Kovarik:2015cma}, and CT14~\cite{Dulat:2015mca}) in order to evaluate part of systematic uncertainties of our calculations. The GRV NLO set~\cite{Gluck:1991ey} has been used for the PDF in a pion (we checked that the SMRS~\cite{Sutton:1991ay} and BSMJ~\cite{Barry:2018ort} sets give almost identical results).
The neutron parton distributions are deduced from those in a proton using isospin symmetry, $f_d^n = f_u^p$, $f_u^n = f_d^p$, $f_{\bar{d}}^n = f_{\bar{u}}^p$, $f_{\bar{u}}^n = f_{\bar{d}}^p$, and $f_{i}^{n} = f_{i}^{p}$ otherwise.

We are interested here in the nuclear dependence of the Drell-Yan process, via the production ratio,
\begin{equation}
    \label{eq:DY_ratio}
    R^{\rm DY}_{\text{h}}(\A/\B,\xf) =  \frac{B}{A}\,\left( \frac{\dd\sigma(\hA)}{\dd \xf} \right) \times  \left( \frac{\dd\sigma(\hB)}{\dd \xf} \right)^{-1},
\end{equation}
in a heavy nucleus (A) over a light nucleus (B). In the absence of nuclear effects, \eq{eq:DY_ratio} may differ from unity because of the differences between proton and neutron parton densities. In practice, proton-induced Drell-Yan collisions probe the target {\it antiquarks} for which $f_{\bar{q}}^n \simeq f_{\bar{q}}^p$, resulting in rather small isospin effects.\footnote{The slight sea quark asymmetry is actually probed from the comparison of DY production in pp and pD collisions. Here, we shall always compare DY yields on nuclei with similar $Z/A$ ratios, making this effect small (see Section~\ref{sec:e906}).} Isospin effects are more important in low energy \piA collisions, mostly sensitive to the valence up and down quark distribution in the nucleus.

\subsection{Nuclear parton distribution functions}\label{subsec:NuclearPDF}

Parton distribution functions in a nucleus differ from those in a free proton over the whole Bjorken-$x$ range, $f_{i}^{p/\A}(x)  \ne  f_{i}^{p}(x)$, where $f_{i}^{p/\A}$ is defined as the PDF of the parton of flavor $i$ inside a proton bound in a nucleus. The latest global fit extractions of nPDF at NLO have been done by DSSZ~\cite{deFlorian:2011fp}, nCTEQ15~\cite{Kovarik:2015cma}, and EPPS16~\cite{Eskola:2016oht} which included for the first time data from the LHC. 

In order to take into account nPDF effects on DY production, $f_{j}^p$ ($f_{j}^n$) needs to be replaced by $f_{j}^{p/\A}$ ($f_{j}^{n/\A}$) in~\eq{eq:PDFnucleus},
\begin{equation}\label{eq:nPDF}
    f_j^{\A}(x_{2}) = Z\,f_{j}^{p/\A}(x_{2}) + (A-Z)\,f_{j}^{n/\A}(x_{2})\,.
\end{equation}
Depending on the parametrizations, either the absolute PDF $f_{j}^{p/\A}$ or the nPDF ratios, $R_{j}^{\A} \equiv f_{j}^{p/\A} / f_{j}^p$, are provided. 

In this article, DY production in \pA and \piA collisions is computed using the nPDF ratios provided by the latest nPDF set, EPPS16~\cite{Eskola:2016oht}. This set actually includes DY data in both \pA and \piA collisions, in addition to other measurements. The implicit assumption is that no other physical effect than a universal leading-twist nuclear PDF would play a role in the production of DY pairs in hadron-nucleus collisions. However, the radiative energy loss of partons may affect the nuclear dependence of DY production, thus spoiling a clean extraction of nPDFs from these data.

\subsection{Initial-state energy loss}\label{subsec:EnergyLoss}

The high-energy partons from the hadron projectile experience multiple scattering while propagating through nuclear matter. This rescattering process induces soft gluon emission, carrying away some of the parton energy available for the hard QCD process, here the Drell-Yan mechanism. The effects of initial-state energy loss on DY can be modelled as~\cite{Arleo:2002ph}
\begin{eqnarray}
    \label{eq:DYxs_eloss}
     \frac{\dd\sigma(h A)}{\dd\xf\,\dd M} &=&   \sum\limits_{i,j=q,\bar{q},g} \int_{0}^{1} \, \dd x_1 \int_{0}^{1} \, \dd x_2\,\int_{0}^{(1-x_1) \Ebeam} \, \dd\epsilon\,\Pdist_i(\epsilon) \, f_{i}^{h}\left(x_{1} + \frac{\epsilon}{\Ebeam}\right) f_{j}^{\A}(x_{2}) \nonumber \\ && \vspace{5cm} \times\, \frac{\dd\widehat{\sigma}_{ij}}{\dd\xf\,\dd M}(x_1 x_2 s)\,,
\end{eqnarray}
where $\Ebeam$ is the hadron beam energy in the rest frame of the nucleus, and $\Pdist_i$ is the probability distribution in the energy loss of the parton $i$~\cite{Baier:2001yt}. The latter has been determined numerically from a Poisson approximation~\cite{Arleo:2002kh,Salgado:2003gb}, using the LPM medium-induced gluon spectrum derived by Baier-Dokshitzer-Mueller-Peigné-Schiff (BDMPS)~\cite{Baier:1996sk}. The first moment of this distribution is given by 
\begin{equation}\label{eq:mean}
    \langle \epsilon_i \rangle \equiv \int_{}^{} \,\dd\epsilon\,\epsilon\, \Pdist_i(\epsilon) = \frac{1}{4}  \alpha_{s} C_{R}\,\qhat\,L^{2}
\end{equation}
where $\alphas=1/2$ is frozen at low scales, $\hat{q} L$ $\lesssim$ 1~GeV$^2$. The medium length $L$ is given by $L=3/4 R$ with $R=r_0\,A^{1/3}$ is the nuclear radius, assuming a hard sphere nuclear density profile ($r_0=(4 \pi \rho / 3)^{-1/3}=1.12$~fm and $\rho$ is the nuclear matter density).

The transport coefficient has been parametrized as (see appendix of~Ref.~\cite{Arleo:2012rs})
\begin{equation}
    \hat{q}(x) \equiv \qzero  \left(\frac{10^{-2}}{x}\right) ^{0.3} \;;\quad x = \min(x_0,x_2)  \;;\quad  x_0 \equiv \frac{1}{2m_p L}\,,
\end{equation}
where the power law behavior reflects the $x$-dependence of the gluon distribution in the nucleus at small values of $x$. In this article we shall consider DY production in low energy hA collisions, probing large values of $x_2$, typically $x_2 > x_0$. Therefore, the transport coefficient used in \eqref{eq:mean} is essentially frozen at $x_0$, $\qhat(x_0) = (0.02\,m_p L)^{0.3}\,\qzero \simeq 0.8\,\qzero$ in a large nucleus ($L=5$~fm). The coefficient $\qzero$ is the only parameter of the model. It has been determined from $\jpsi$ data using the energy loss in the fully coherent regime, $\qzero = 0.07$--$0.09$~\gevsqfm~\cite{Arleo:2012rs}. No attempt has been made here to extract an independent estimate of $\qhat$ from low energy DY data, as the measurements from the E906 experiment are still preliminary. Eventually, this would allow one to check the universality of this parameter, for different processes and in different energy loss regimes (LPM for DY, fully coherent for $\jpsi$; see Introduction). 

The BDMPS formalism is particularly suited to describe gluon radiation induced by multiple soft scattering, thus appropriate for thick media for which the typical number of scatterings, $n=L/\lambda$, is large ($\lambda$ is the parton mean free path). This is at variance with the GLV~\cite{Gyulassy:1999zd} and higher-twist approach~\cite{Wang:2001if}, which instead consider a single hard scattering in a medium. Using $\qhat=\mu^2/\lambda$, where $\mu\simeq 200$~MeV is the typical momentum transfer in single soft scattering, the number of scatterings in a big nucleus (taking $L=5$~fm) is $n=\qhat L / \mu^2 =7$--$9$, which is consistent with the initial assumption of BDMPS.\footnote{Note that in a single hard picture, the momentum transfer neeeds to be large enough in order to induce gluon emission. Taking $\mu_{\text{semi-hard}}\simeq 500$~MeV leads to $n\simeq1$, also consistent with the assumption of a small number of (semi-hard) scattering.}

As can be seen from \eq{eq:DYxs_eloss}, the nuclear dependence of DY production depends on the shape of the hadron beam PDF: the steeper the $x$-dependence of $f_i^h$, the stronger the DY suppression. Since at large $x$ one expects  $f_q^p(x)\sim(1-x)^3$ and $f_q^\pi(x)\sim(1-x)^2$ from quark counting rules~\cite{Brodsky:1973kr}, a stronger DY suppression can be expected in \pA collisions with respect to that in \piA collisions~\cite{Arleo:2002ph}. Another consequence follows from \eq{eq:DYxs_eloss}. At large \xf, the maximal amount of parton energy loss is restricted to be $\epsilon < (1-\xone)\,\Ebeam \simeq (1-\xf)\,\Ebeam$, making DY production dramatically suppressed at the edge of phase-space, $\xf \lesssim 1$.

\section{Phenomenology}\label{sec:phenomenology}

On top of isospin effects (labelled CT14 in the figures), the DY nuclear production ratio \eqref{eq:DY_ratio} is computed assuming either nPDF effects, as estimated using the EPPS16 parton densities in \eq{eq:PDFnucleus} and their associated error sets, or initial-state energy loss effects, \eq{eq:DYxs_eloss}. Although each effect is here studied separately, both could in principle be taken into account in order to achieve a complete description of the Drell-Yan process in hA collisions.

The calculations are compared to the preliminary results from E906\footnote{These are not included in the global fit of EPPS16.} at $E_{p}=120$~GeV~\cite{Lin:2017eoc}, NA10 data at $E_{\pi^-}=140$~GeV~\cite{Bordalo:1987cs} and E866 at $E_{p}=800$~GeV~\cite{Vasilev:1999fa}. Predictions for the COMPASS experiment, which are being collected Drell-Yan data in \piA collisions at $E_{\pi^-}$=190~GeV~\cite{Aghasyan:2017jop}, are also presented.

\subsection{E906 preliminary data}\label{sec:e906}

\begin{figure}[ht]
   \begin{minipage}[c]{.44\linewidth}
      \includegraphics[scale=0.4]{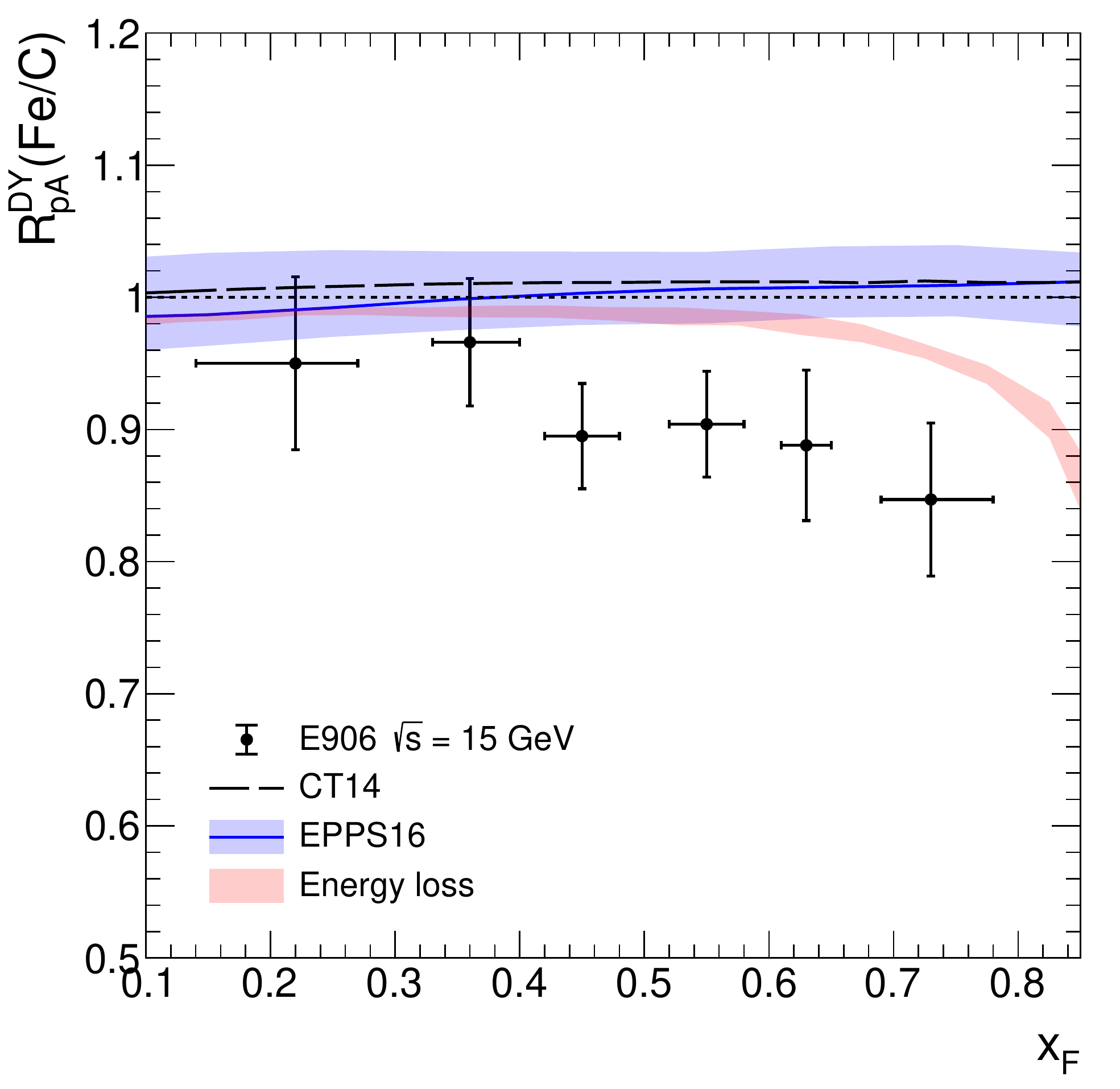}
   \end{minipage} \hfill
   \begin{minipage}[c]{.44\linewidth}
      \includegraphics[scale=0.4]{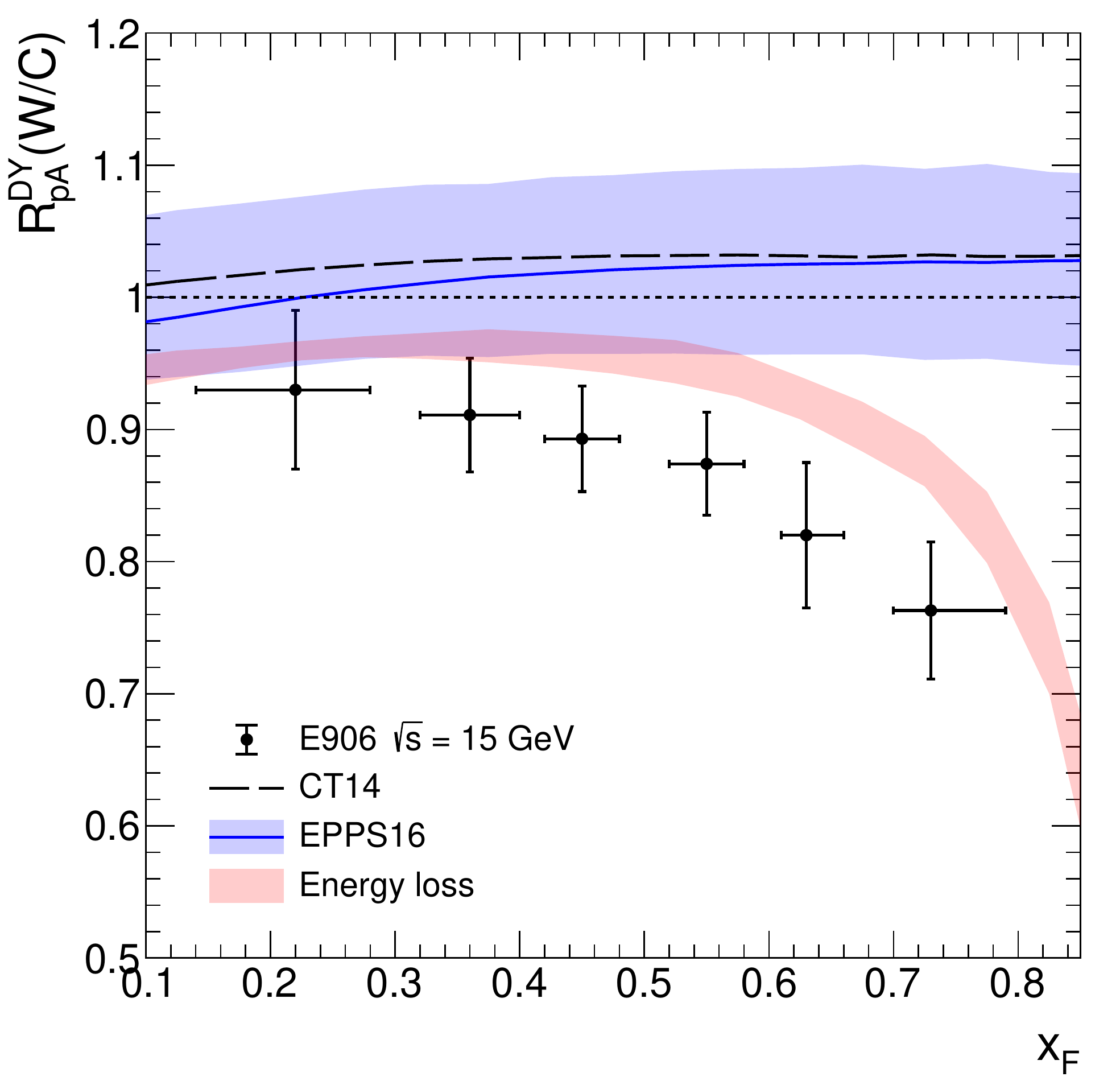}
   \end{minipage}
   \caption{E906 nuclear production ratio measured in pFe (left) and pW (right), normalized to pC, collisions at $\sqrts=15$~GeV compared to EPPS16 nPDF calculation (blue band), isospin effect (dotted line) and energy loss effects (red band).}
   \label{E906_data}
\end{figure}

The preliminary results from the E906 experiment~\cite{E906,Lin:2017eoc} on the ratios $\rpa(\text{Fe/C})$ and $\rpa(\text{W/C})$ are shown as a function of \xf in Fig.~\ref{E906_data}. The mass range is $4.5 < M < 5.5$~GeV  at $\sqrts = 15$~GeV with an additional kinematical cut, $0.1 < x_{2} < 0.3$. 

The data indicate a clear DY suppression in both Fe and W nuclear targets, increasingly pronounced at large \xf, in clear contrast with the nPDF calculations shown as a blue band. In particular, nPDF effects make \rpa consistent with the CT14 predictions assuming no nuclear modification of parton densities, see \eqref{eq:DYxs}, shown as a dashed line in Fig.~\ref{E906_data}. This \xf range indeed corresponds to typical values of $\xtwo\simeq 0.1$--$0.3$, at the boundary between the EMC effect and the antishadowing region, hence for which $f_i^{p/\A}\simeq f_i^p$. The slight rise of \rpa with increasing \xf, seen for both CT14 and EPPS16, originates from the asymmetry in the nucleon sea, $f_{\bar{d}}^p > f_{\bar{u}}^p$.\footnote{Thus in a W target, for which neutrons are more abundant than protons, one expects $f_{\bar{u}}^{\rm W} > A_{\rm W}\,f_{\bar{u}}^p$, leading to $\rpa$ slightly above unity.} To our knowledge, this is the first time that DY measurements in hA collisions exhibit such a clear disagreement with the nPDF expectations.

The initial-state energy loss effects, assuming the default choice of $\qzero=0.07$--$0.09$~\gevsqfm, are shown as a red band. The model uncertainty includes the choice of different proton PDF sets, on top of the variation of $\qzero$. The calculations predict a significant DY suppression which becomes more pronounced as \xf gets larger, in qualitative agreement with the E906 results. While the magnitude of \rpa is smaller in the measurements than in the model, by roughly 5\% and 10\% in W and Fe targets respectively, we find it nevertheless remarkable that the \emph{shapes} of \rpa predicted by the energy loss model and the E906 results prove similar.\footnote{In particular, the calculation of \rpa does not depend on any free parameter.} In our opinion, these preliminary data strongly hint at the existence of (LPM) initial-state quark energy loss in the DY process.
The origin of the different magnitude in the data and in the model is not clear. We checked that a larger transport coefficient would reproduce perfectly the W/C ratio, however at the expense of a non-universal coefficient in the LPM and fully coherent energy loss regimes. Also, note that the present discrepancy is of the same order as the nPDF uncertainty; adding  initial-state energy loss together with nPDF effects would thus come in good agreement with the data. Finally, it cannot be excluded that fully coherent energy loss, expected at NLO in the $q\bar{q}\to \gamma^\star g$, could lead to an extra suppression at large \xf. Any conclusion nonetheless awaits for the E906 measurements to become final.

Earlier calculations have been performed in a different theoretical set-up, the higher twist (HT) formalism, which allows for computing within the same framework the DY production process and the energy loss effects, assuming one additional hard scattering in the target nucleus~\cite{Xing:2011fb}. These calculations should match at leading twist (assuming no additional scattering) the leading order DY production cross section. On the contrary, the DY process is here computed at NLO accuracy to which a rescaling of the projectile is PDF is assumed, in order to model energy loss effects arising from multiple soft scattering in the medium, see \eq{eq:DYxs_eloss}. Despite these important differences, both calculations of the DY nuclear production ratio prove remarkably similar (and the HT calculation of Ref.~\cite{Xing:2011fb} in agreement with E906 preliminary data). It should also be mentioned that the transport coefficient used in both approaches coincide, as the \emph{gluon} transport coefficient used here would correspond to a \emph{quark} transport coefficient given by $\qhat_{\text{quark}} = 4/9\, \qhat \simeq 0.025$--$0.032$~\gevsqfm (in a W nucleus) perfectly matches the value $\qhat_{\text{quark}}=0.024\pm0.008$~\gevsqfm used in Ref.~\cite{Xing:2011fb} and extracted from semi-inclusive DIS measurements~\cite{Wang:2009qb}.

\subsection{NA10 data}

The NA10 collaboration collected Drell-Yan data on two nuclear targets (W, D) and in the mass range is $4.35 < M < 15$~GeV  at $\sqrts = 16.2$~GeV, excluding the $\Upsilon$ peak region $8.5< M < 11$~GeV. 

\begin{figure}[bt!]
\begin{center}
\includegraphics[scale=0.4]{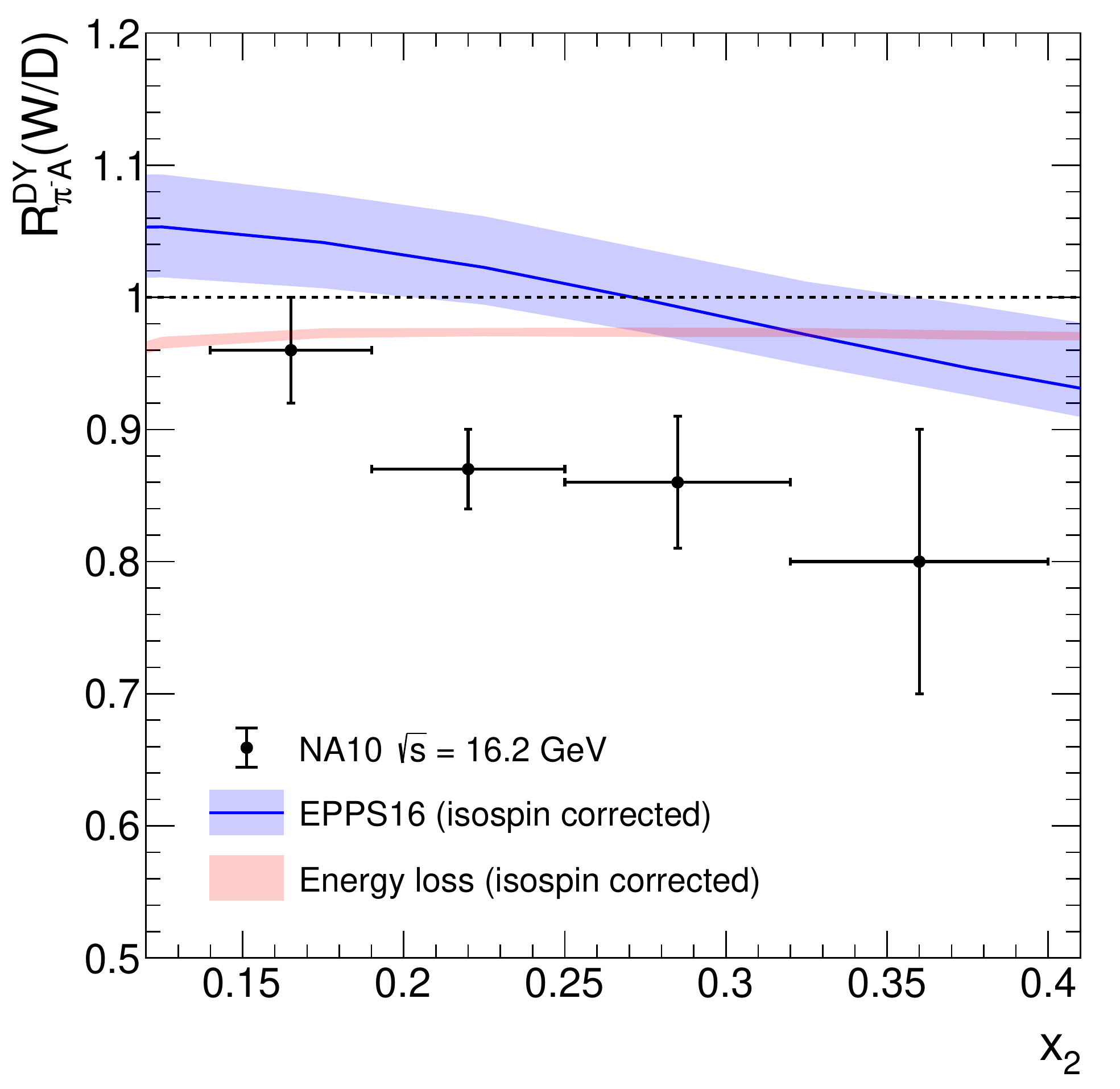}
\end{center}
\caption{NA10 nuclear production ratio, corrected for isospin effects, measured in $\pi$W, normalized
to $\pi$D, collisions at $\sqrts=16.2$~GeV compared to EPPS16 nPDF calculation (blue band) and energy loss effects (red band).}
   \label{NA10_data}
\end{figure}

The original measurements were corrected for isospin effects in the W target~\cite{Bordalo:1987cs}. Similarly, an isospin correction~\cite{Paakkinen:2016wxk} is applied to the present calculation, defined as
\begin{equation}
    R_{\pi^-}^{\text{NLO-isospin corrected}}(\text{W/D}) = R_{\pi^-}^{\text{LO}}(\text{W}^{\text{isoscalar}}\text{/W})_{\text{no nPDF}} \times R_{\pi^-}^{\text{NLO}}(\text{W/D}),
\end{equation}
where $R_{\pi^-}^{\text{LO}}(\text{W}^{\text{isoscalar}}\text{/W})_{\text{no nPDF}}$ is calculated at leading order in an 
`isospin-symmetrized' W nucleus ($Z=A/2$) over that in a W nucleus. The isospin corrected nPDF calculation overestimates the data, as illustrated in Fig.~\ref{NA10_data}. This disagreement has also been reported by Paakkinen et al.~\cite{Paakkinen:2016wxk} who apply a rescaling factor of $12.5\%$ ($r=1.125$) in their calculations to make data and nPDF corrections come in agreement. This rescaling  is however twice larger than the systematic uncertainties of $6\%$ reported in the experiment~\cite{Bordalo:1987cs}.

The energy loss calculation shown in Fig.~\ref{NA10_data} leads to a suppression of approximately $3\%$ ($\rpa\simeq0.97$), independent of $x_2$. As expected, energy loss effects in \piA collisions turn out to be significantly smaller than in \pA collisions due to the harder quark distribution in a pion with respect to that in a proton (see Section~\ref{subsec:EnergyLoss}). Taking into account energy loss in addition to nPDF effects would thus require a rescaling factor of approximately $1.125 \times 0.97 = 9\%$, instead of $12.5\%$ previously, hence closer to the reported experimental systematic uncertainty.

Perhaps more importantly, the present calculation shows that initial-state energy loss gives an effect on \rpia of the same magnitude as that of nPDF corrections. This may thus question a reliable extraction of nuclear parton distributions from DY data in \piA collisions without taking energy loss effects into account. Moreover, although \rpia proved to be constant (but still different from unity) in this $\xtwo$ range, we shall see in Section~\ref{sec:compass} that it is not necessarily always the case.

\subsection{E866 data}

The E866 collaboration measured the Fe/Be and W/Be nuclear production ratios in \pA collisions at $\sqrts = 38.7$~GeV. The integrated mass range is $4 < M < 8$~GeV, with an additional kinematical cut $0.02 \lesssim x_{2} \lesssim 0.10$. 

The nuclear depletion of DY production in E772 and E866 data has long been delicate to interpret as it is virtually impossible to disentangle both nPDF and energy loss processes from these sole measurements~\cite{Arleo:2002ph}. Several groups attribute the measurements as coming mostly from large energy loss effects~\cite{Johnson:2001xfa,Neufeld:2010dz,Song:2017wuh}, while the nPDF analyses instead assume energy loss effects to be negligible and hence do incorporate these experimental data in the global fits~\cite{deFlorian:2011fp,Kovarik:2015cma,Eskola:2016oht}. 
\begin{figure}[ht!]
   \begin{minipage}[c]{.46\linewidth}
      \includegraphics[scale=0.4]{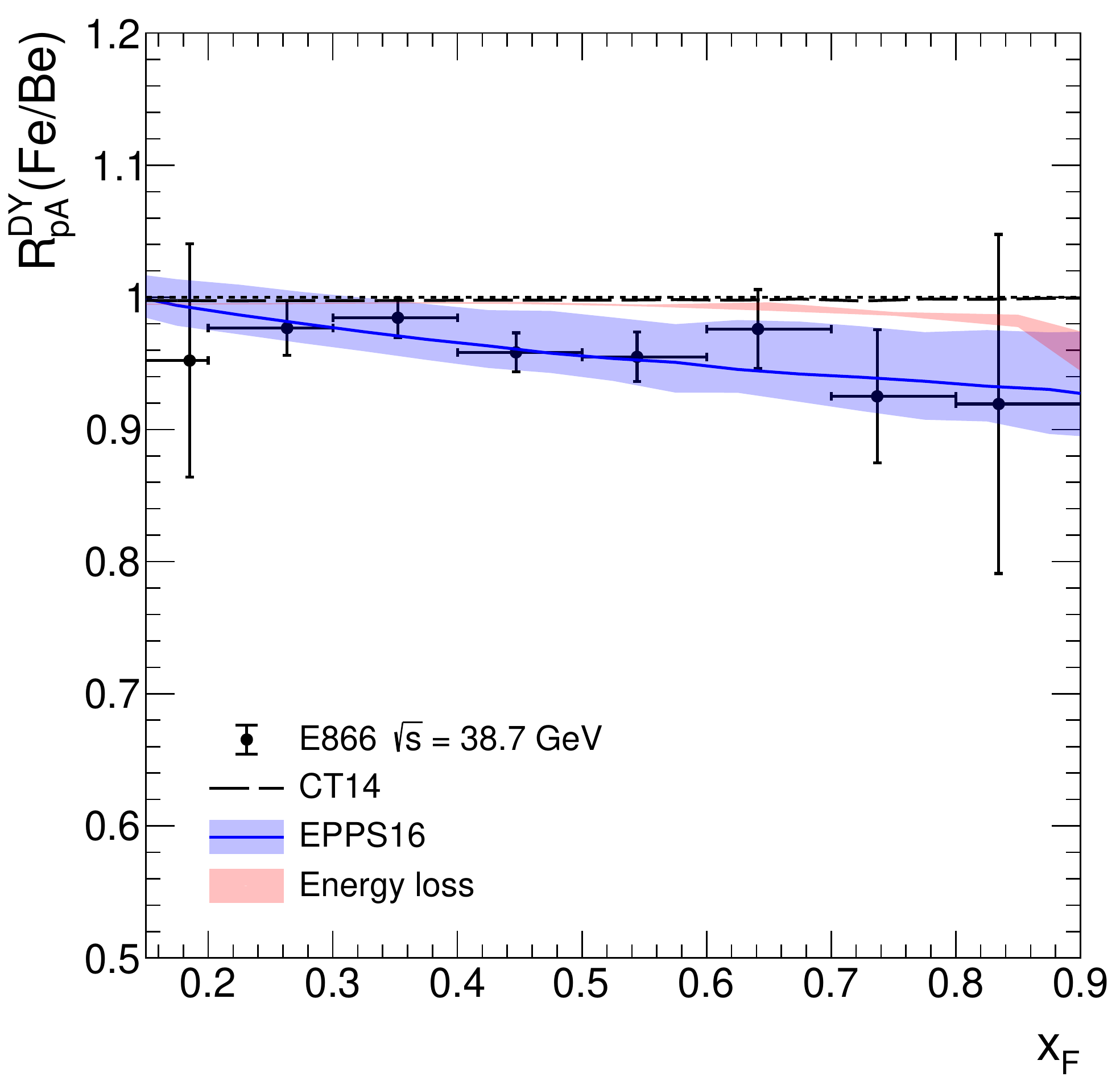}
   \end{minipage} \hfill
   \begin{minipage}[c]{.46\linewidth}
      \includegraphics[scale=0.4]{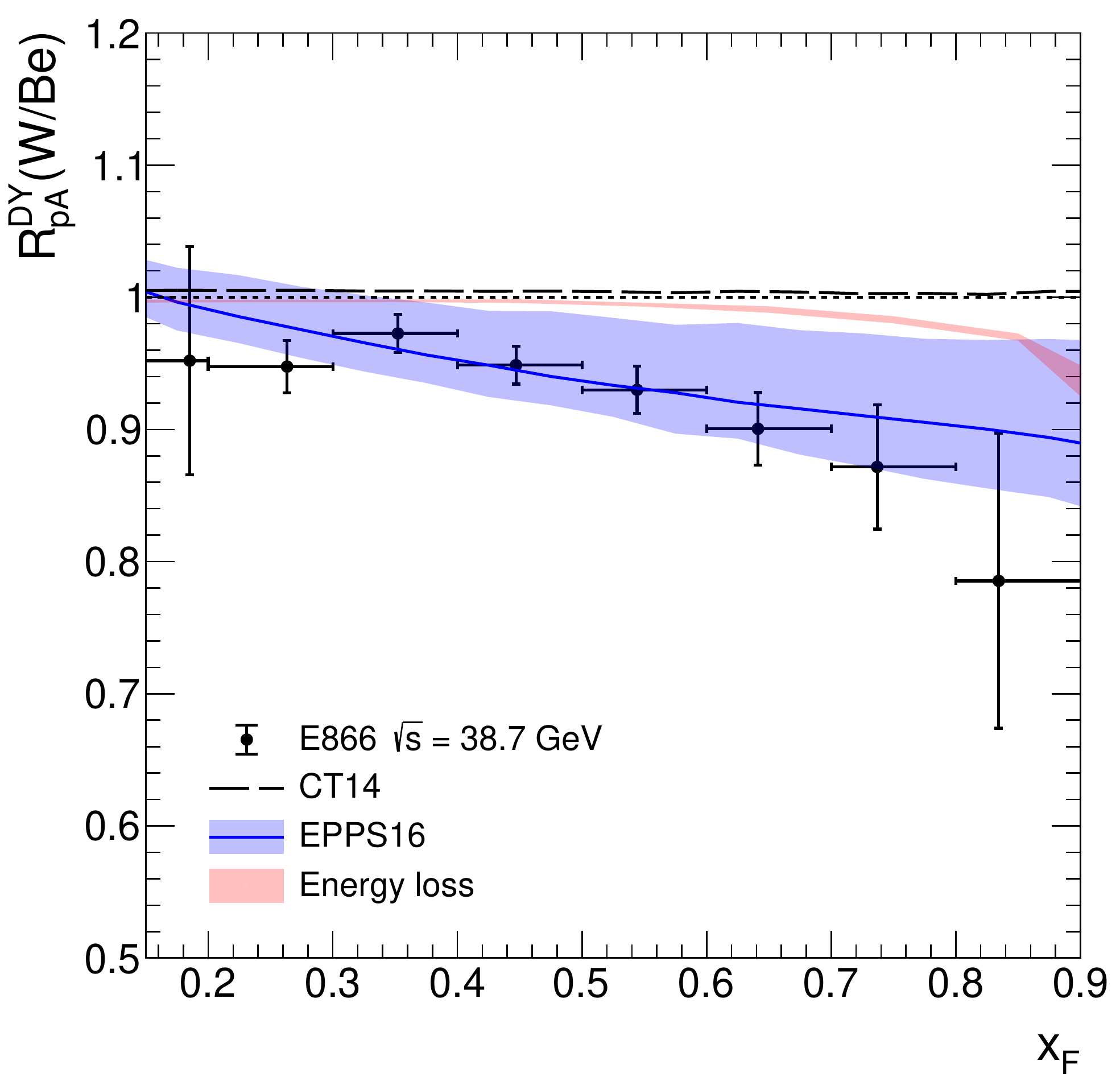}
   \end{minipage}
   \caption{E866 nuclear production ratio measured in pFe (left) and pW (right), normalized to pBe, collisions at $\sqrts=38.7$~GeV compared to EPPS16 nPDF calculation (blue band), isospin effect (dotted line) and energy loss effects (red band).}
   \label{E866_data}
\end{figure}
The comparison between our results and the data is shown in Fig.~\ref{E866_data}. The agreement between nPDF calculation and the data is satisfactory for both nuclear ratios. However, this should not come as a surprise since these data have been used in the global fit of EPPS16.
In Ref.~\cite{Arleo:2002ph}, NA3 DY data in \piA collisions were used at lower energy to extract an upper limit on the amount of quark energy loss in nuclear matter, and hence helped to lift this `degeneracy' at E866 energy. That study led to the conclusion that energy loss effects on the E866 DY measurements were presumably small.
Here, the independent estimate of $\qhat$ from \jpsi suppression data --~which are not included in nPDF analyses~-- corroborates this statement as parton energy loss shows a negligible effect, except at very large $\xf \gtrsim 0.8$ (see Fig.~\ref{E866_data}). At E866 energy and above, the forward DY measurements in \pA collisions are therefore most likely due to nPDF effects and hence could be included in the global fit analyses.\footnote{The extraction of sea quark nPDF at small $\xtwo\sim10^{-2}$ (corresponding the largest \xf bin in the experiment) is probably affected by a few percent.} At LHC, the coming DY measurements in pPb collisions at the LHC should thus allow for a clean extraction of nPDF at small $x$~\cite{Arleo:2015qiv}.

\subsection{Predictions for the COMPASS experiment}\label{sec:compass}

Drell-Yan data are also being collected by the COMPASS collaboration at the CERN SPS, using a pion beam on two nuclear targets (NH$_{3}$, W) at a collision energy $\sqrts=18.9$~GeV~\cite{Aghasyan:2017jop}. The expected mass range is $4.3 < M < 8.5$~GeV. With such a mass range, the COMPASS measurements would explore a typical range in $0.1\lesssim\xtwo\lesssim0.5$, embracing both the antishadowing and EMC regions.

\begin{figure}[t!]
\begin{center}
 \includegraphics[scale=0.4]{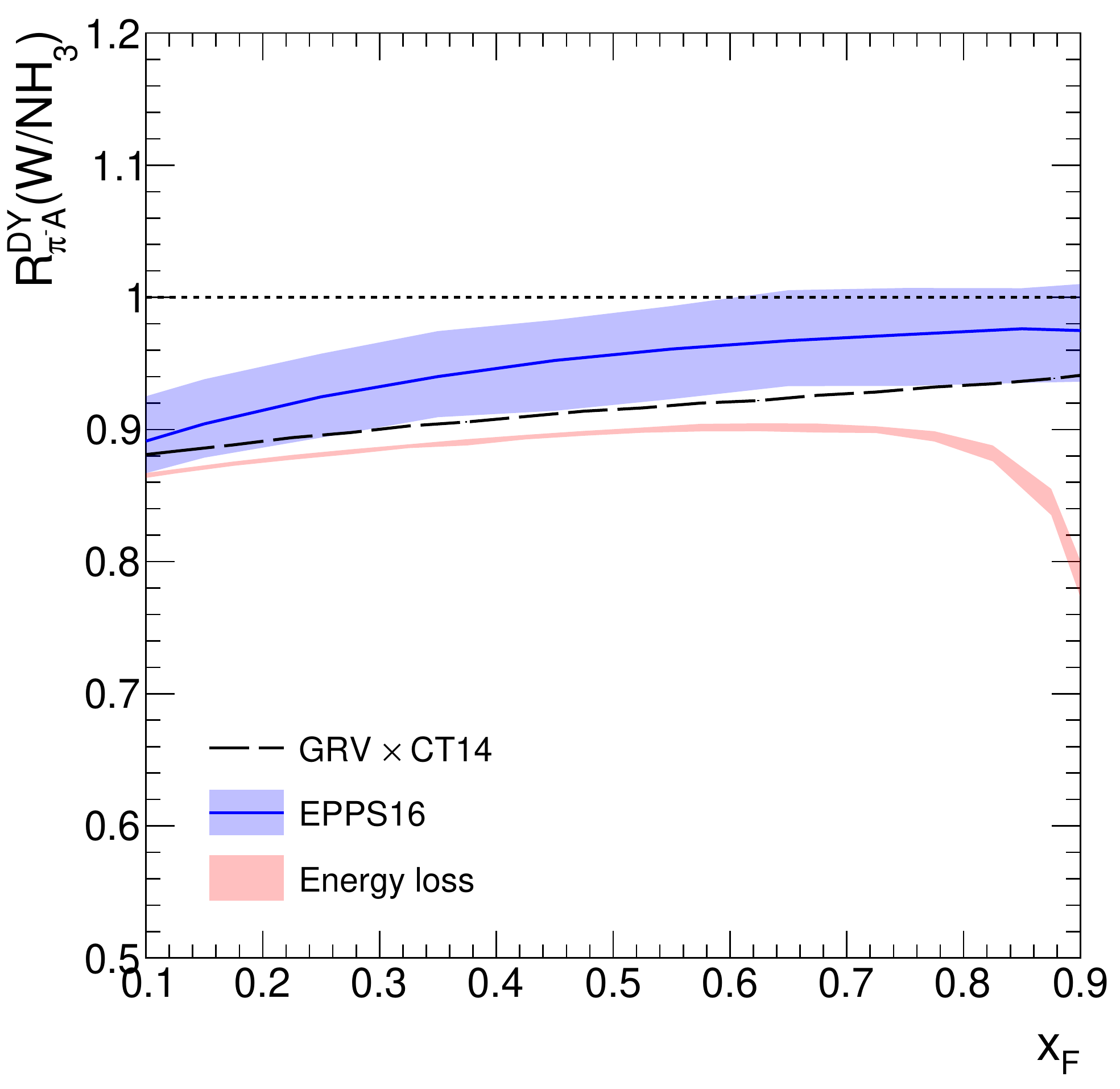}
 \caption{Nuclear production ratio measured in $\pi$W, normalized to $\pi$NH$_3$, collisions  at $\sqrts=18.9$~GeV compared to EPPS16 nPDF calculation (blue band), isospin effect (dotted line) and energy loss effects (red band)}
 \label{COMPASS_data}
 \end{center}
\end{figure}

The predictions for the ratio $R_{\pi^{-}\text{A}}^{\text{DY}}$(W/NH$_{3}$) are shown in Fig.~\ref{COMPASS_data}. A significant suppression from isospin effects (dashed curve) are expected because of the lesser up quark density in W than in NH$_3$ nuclear target.\footnote{Using $Z_{\text{W}}/A_\text{W}\approx2/5\simeq 1-Z_{\text{NH}_3}/A_{\text{NH}_3}$ leads to an expected suppression $\rpia\simeq (1+3/2\,r)/(3/2+r)\approx7/8$ assuming $r\equiv d/u\approx1/2$ at large \xtwo. At smaller \xtwo, hence at larger \xf, $r$ is getting larger and so does $\rpia$.}
Because of the valence quark antishadowing in the EPPS16 set, the inclusion of nPDF corrections increases $\rpia$ by up to $5\%$ with respect to the calculation assuming no nuclear effect.\footnote{The EMC region is probed at smaller values of \xf, not shown in Fig.~\ref{COMPASS_data}.}

On the contrary, energy loss effects would lead to a suppression of the DY yield, increasingly pronounced at larger \xf as the phase space available for gluon radiation shrinks. At $\xf=0.9$, the model predicts a suppression of $\rpia\simeq0.8$ while isospin effects only would lead to $\rpia\simeq0.94$. Despite the smaller energy loss effects in \piA collisions, the future DY measurements by the COMPASS experiment will provide additional constraints for the extraction of the transport coefficient, thanks to the large \xf acceptance and the expected statistics.

Before closing this section, let us mention that the pion PDF are extracted from DY production measured in pion-induced collisions on \emph{nuclear} targets, assuming no nuclear effect beyond isospin corrections. However, energy loss effects may affect the reliable extraction of the pion PDF at large $x$, $f_u^\pi(x)\sim(1-x)^n$, and make $n$ possibly overestimated in the PDF global fits. In order to estimate the error associated to the use of nuclear targets, the DY nuclear production ratio has been fitted as $\rpia(\xone) = (1-\xone)^{\delta n}$, where $\delta n \simeq 0.06$ is a typical correction to the slope of the pion PDF.

\section{Violation of factorization in DY production in pA collisions}\label{sec:x2scaling}

\subsection{Factorization and \texorpdfstring{$x_2$}{x2} scaling}

It has been shown in the previous section that calculations using EPPS16 nPDF sets fail to describe the preliminary results by the E906 experiment (Fig.~\ref{E906_data}), while exhibiting a good agreement with E866 measurements (Fig.~\ref{E866_data}). It could be argued that \emph{this} specific nPDF set is unable to account for both data sets. On the contrary, we demonstrate here that the failure to describe both data sets should be generic to all calculations based on nPDF effects only, following the reasoning of Ref.~\cite{Hoyer:1990us}.

Let us consider the factorized expression for the DY production cross section in \pA collisions, \eq{eq:DYxs}, assuming possible nPDF corrections, \eq{eq:nPDF}. In the forward region, $\xf = \xone - \xtwo >0$, and at large collision energy ($\hat{s}/s\simeq M^2/s\ll 1$), the momentum fractions carried by the incoming partons can be approximated as $\xone \simeq \xf$ and $\xtwo = \hat{s} / (\xf s)$. Since forward DY production is dominated by the scattering of a quark in the incoming proton, the \pA differential cross section \eqref{eq:DYxs} can thus be approximated as~\cite{Hoyer:1990us}
\begin{eqnarray}
    \label{eq:DYxs_approx}
     \frac{\dd\sigma(\pA)}{\dd\xf\,\dd M} &\simeq& f_{q}^{p}(\xf) \times \left( \sum\limits_{j=q,\bar{q},g} \, \int\, \dd\xtwo\, f_{j}^{\A}(x_{2}) \frac{\dd\widehat{\sigma}_{qj}}{\dd\xf\,\dd M}(\xf \xtwo s)\right)\, \nonumber \\
     &\simeq& f_{q}^{p}(\xf) \times \left( \sum\limits_{j=q,\bar{q},g} f_{j}^{\A}(x_{2})\, \frac{\dd\widehat{\sigma}_{qj}}{\dd\xf\,\dd M}(M^2)\right)\,
\end{eqnarray}
where the second line is obtained assuming that the partonic cross section peaks close to the threshold, $\hat{s} \gtrsim M^2$.

Using \eq{eq:DYxs_approx}, the nuclear production ratio \eqref{eq:DY_ratio} thus becomes a scaling function of the \xtwo momentum fraction only --~independent of the center-of-mass energy of the collision~-- should the factorization hold in DY forward production in \pA collisions. Conversely, a lack of $\xtwo$ scaling in data would signal the breakdown of QCD factorization and would indicate that nPDF corrections alone cannot account for the nuclear dependence of DY production.

\subsection{\texorpdfstring{$x_2$}{x2} scaling violation in the DY process}

\begin{figure}[ht!]
   \begin{minipage}[c]{.46\linewidth}
      \includegraphics[scale=0.4]{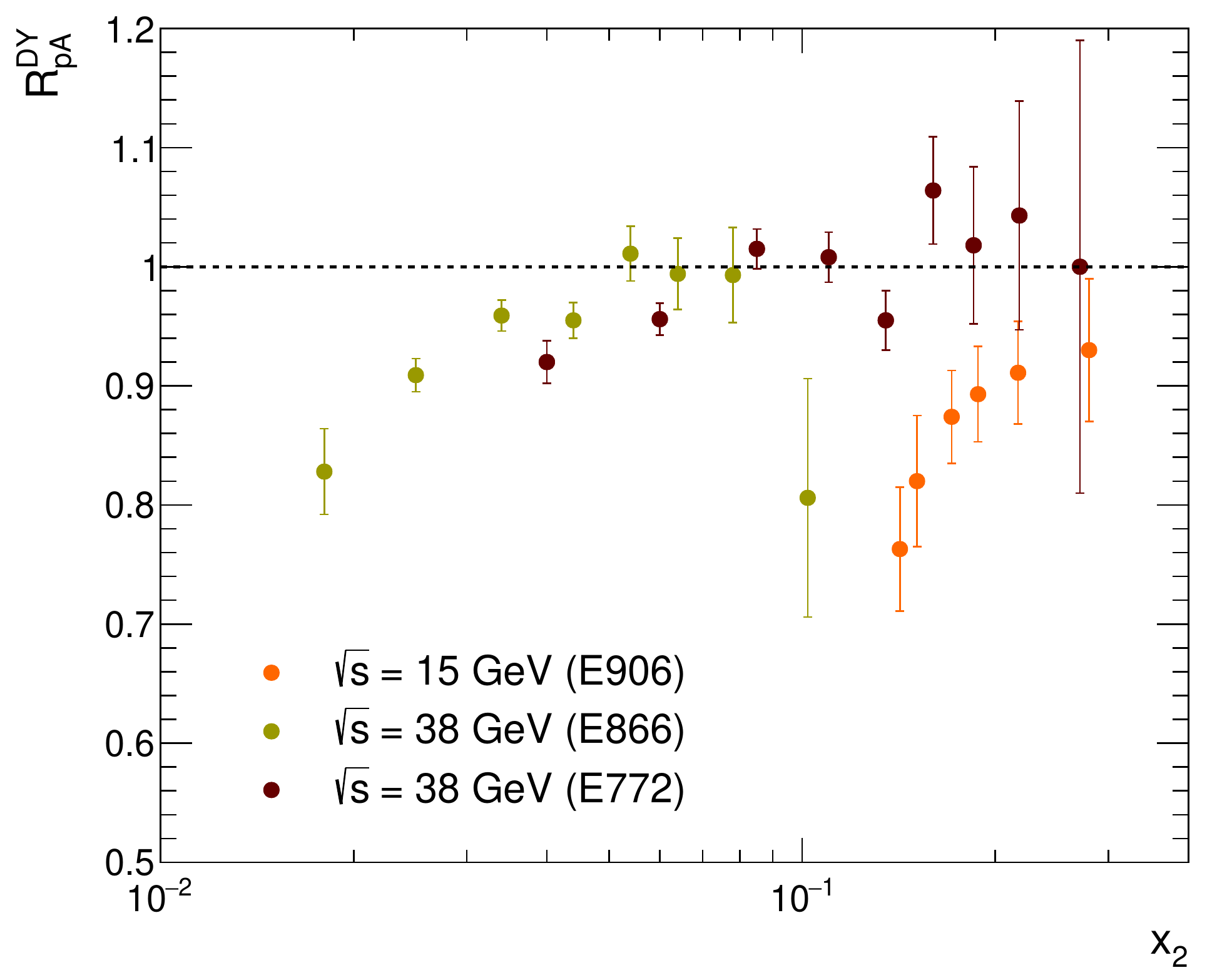}
   \end{minipage} \hfill
   \begin{minipage}[c]{.46\linewidth}
     \includegraphics[scale=0.4]{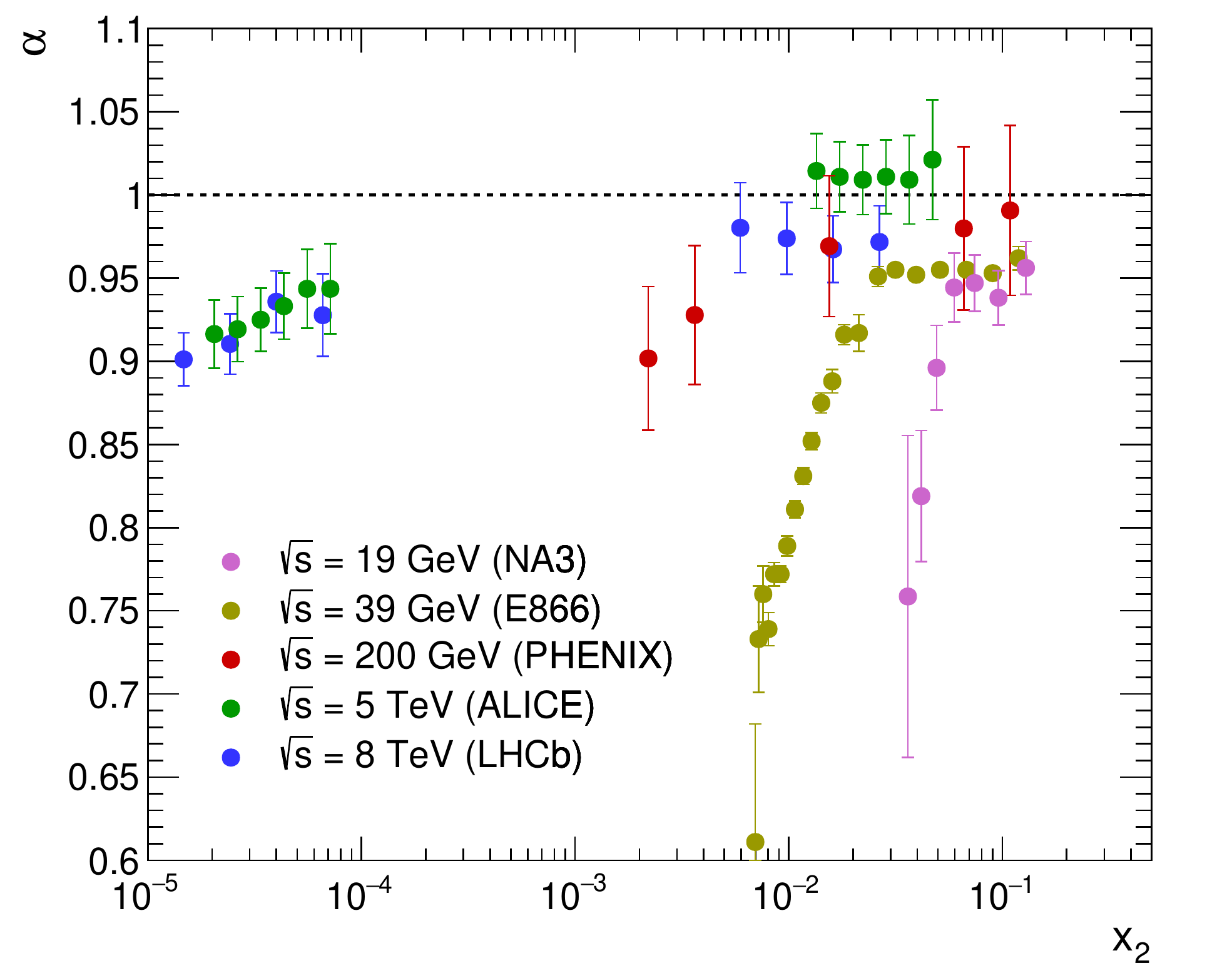}
   \end{minipage}
    \caption{Left: DY nuclear production ratio measured by E772~\cite{Alde:1990im}, E866~\cite{Vasilev:1999fa}, E906~\cite{Lin:2017eoc}, plotted as a function of \xtwo.
    Right: Nuclear dependence of $\jpsi$ production measured by NA3~\cite{Badier:1983dg}, E866~\cite{Leitch:1999ea}, PHENIX~\cite{Adare:2010fn}, ALICE~\cite{Abelev:2013yxa} and LHCb~\cite{Aaij:2017cqq}, plotted as a function of \xtwo.}
   \label{fig:x2scaling}
\end{figure}

In order to check whether the DY nuclear production ratio indeed scales like \xtwo, the \pA data from E772 (on W/D targets), E866 (both taken at $\sqrts=38.7$~GeV) and E906 ($\sqrts=15$~GeV) are plotted as a function of \xtwo in Fig.~\ref{fig:x2scaling} (left). The comparison between E772 and E906 results clearly show a violation of $\xtwo$ scaling at large $\xtwo\sim 10^{-1}$.\footnote{In principle the backward ($-2.2 < y < -1.2$) DY data by PHENIX in pAu collisions at $\sqrts=200$~GeV could probe similar values of $\xtwo$, yet the present uncertainties prevent any quantitative conclusions yet~\cite{phenixdy}. Similarly, future measurements by the LHCb experiment in pPb collisions at $\sqrts=8.16$~TeV~\cite{Bediaga:2018lhg} may also access the large $\xtwo$ domain.} On the contrary, it has been checked that the NLO calculations (using EPPS16 nPDF sets) at both energies follow, as expected, $\xtwo$~scaling to a very good accuracy.

This comparison thus provides for the first time a clear evidence of the violation of QCD factorization in the Drell-Yan process in \pA collisions, implying the presence of higher-twist processes. A natural candidate responsible for the reported violation of $\xtwo$ scaling is LPM initial-state energy loss, as the comparison between the model calculation and E906 results suggests. As mentioned in the Introduction, initial-state energy loss should have a weak impact at high incoming parton energy, $E = \xone \Ebeam = M^2/(\xtwo s)$. Therefore, no violation of $x_2$ scaling in DY production is expected at small values of $\xtwo$, as $\meaneps_{\text{LPM}}/E \propto \xtwo$.

\subsection{Comparing with \texorpdfstring{$\jpsi$}{J/psi} production}

Such \xtwo scaling breakdown has long been reported in \jpsi production~\cite{Hoyer:1990us}, later confirmed by measurements at higher collision energy~\cite{Leitch:2006ff}. Fig.~\ref{fig:x2scaling} (right) shows the nuclear dependence of $\jpsi$ production in \pA collisions\footnote{parametrized by $\alpha$, where $\alpha$ is defined as $\sigma(\pA\to \jpsi\,\X) \equiv \sigma(\text{pp}\to\jpsi\,\X)\times A^\alpha$} measured by NA3~\cite{Badier:1983dg}, E866~\cite{Leitch:1999ea}, PHENIX~\cite{Adare:2010fn}, ALICE~\cite{Abelev:2013yxa} and LHCb~\cite{Aaij:2017cqq}, from $\sqrts\simeq 20$~GeV to $\sqrts\simeq 8$~TeV. As can be seen, the $\jpsi$ suppression data clearly rule out the $\xtwo$ scaling predicted by QCD factorization. This prevents the use of $\jpsi$ measurements in \pA collisions in order to extract nuclear parton densities.

The $\xtwo$ dependence of DY and $\jpsi$ suppression at different collision energies show similar patterns, despite a clear difference in the magnitude of the suppression.\footnote{The suppression is much more pronounced in the $\jpsi$ channel than in the DY process. Taking \eg $\alpha=0.75$ would lead to $R^{J/\psi}_{pA} \simeq 0.27$ in a $A=200$ nucleus.} In the case of \jpsi production, the good agreement between data and the model based on fully coherent energy loss~\cite{Arleo:2012hn,Arleo:2012rs} makes the latter a natural process responsible for the breakdown of $\xtwo$ scaling. Unlike initial-state energy loss, fully coherent energy loss is proportional the parton energy, \eq{eq:coherent}, making $\meaneps_{\text{coh}}/E$ independent of $\xtwo$. As a consequence, quarkonium suppression should not follow $\xtwo$ scaling even at small values of $\xtwo$.

\section{Summary}\label{sec:summary}

The effects of parton energy loss on Drell-Yan production in \pA and \piA collisions at fixed-target energies has been investigated, based on a BDMPS energy loss framework embedded in a NLO calculation. Nuclear production ratios were compared to calculations assuming nPDF effects as well as to experimental results. Let us summarize the main conclusions of this study:
\begin{itemize}
    \item Preliminary results by the E906 experiment in \pA collisions exhibit a significant DY suppression at large \xf, in clear disagreement with calculations including nPDF corrections only. This is the first time that the nuclear dependence of the DY process contradicts unambiguously nPDF expectations, indicating that other processes should be at work. This conclusion is confirmed by the direct comparison of E906 and E772/E866 results, which lack of $\xtwo$ scaling signals the violation of QCD factorization in Drell-Yan production in \pA collisions.
    \item In contrast, the E906 results prove in good qualitative agreement with the energy loss model predictions, despite a slightly different magnitude, using the transport coefficient extracted from \jpsi data. This is a clear hint that energy loss in cold QCD matter affects the Drell-Yan process in nuclear collisions, while earlier claims of energy loss effects on DY production at higher collision energy are spoiled by possible nPDF effects. Moreover, the qualitative agreement with E906 preliminary data points to a consistent value of the transport coefficient for two processes in two different dynamical regimes: LPM energy loss for Drell-Yan and fully coherent radiation for \jpsi.
    \item In \piA collisions, energy loss effects are naturally smaller than in \pA collisions due to the harder PDF in a pion with respect to that in a proton. However, it is shown that energy loss effects are of the same magnitude as nPDF corrections, sowing seeds of doubt on a clean extraction of nPDF from these data without including energy loss effects. Energy loss processes suppress the DY yield and helps reducing the known tension between nPDF results and NA10 data, however not sufficiently to claim for a good agreement. Predictions for the future COMPASS results are made; significant energy loss effects are expected especially above $\xf \gtrsim 0.7$ which should bring additional constraints on the cold nuclear matter transport coefficient.
    \item At E866 energy ($\sqrt{s} = 38.7$ GeV) the effects of LPM energy loss on DY production, using $\qhat$ extracted from \jpsi data, significantly weaken, as already pointed out in Refs.~\cite{Arleo:2002ph,Xing:2011fb}. This justifies a posteriori the use of these results to extract nPDF, except at very large \xf where energy loss affects DY almost as much as nPDF corrections.
\end{itemize}

\acknowledgments

We would like to thank Yann Bedfer, Fabienne Kunne and Stéphane Peigné for a careful reading of the manuscript and for discussions. We also thank Po-Ju Lin for discussions on the E906 results. This work was supported in part by the P2IO LabEx (ANR-10-LABX-0038).

\providecommand{\href}[2]{#2}\begingroup\raggedright\endgroup
\end{document}